\documentclass[showpacs,twocolumn,prl]{revtex4-1}
\usepackage[utf8]{inputenc}
\usepackage{graphicx}
\usepackage{color}
\usepackage{appendix}
\usepackage{amsmath}
\usepackage{amssymb}
\usepackage{subfigure}
\usepackage{epsfig}

\newcommand{\kapp}{\boldsymbol{\kappa}}

\newcommand{\om}{\hat{\Omega}}

\newcommand{\omadj}{{\Omega}^{\dagger}}
\newcommand{\F}{\boldsymbol{F}}

\newcommand{\kk}{\boldsymbol{k}}
\newcommand{\qq}{\boldsymbol{q}}

\newcommand{\gam}{\dot{\gamma}}

\newcommand{\p}{\boldsymbol{p}}
\newcommand{\rr}{\boldsymbol{r}}
\newcommand{\nab}{\boldsymbol{\nabla}}
\newcommand{\xxi}{\boldsymbol{\xi}}
\newcommand{\eeta}{\boldsymbol{\eta}}
\newcommand{\R}{\boldsymbol{R}}
\newcommand{\Phik}{\Phi_{\boldsymbol{k}}}
\newcommand{\Gammak}{\Gamma_{\boldsymbol{k}}}
\newcommand{\mk}{m_{\boldsymbol{k}}}

\newcommand{\rhok}{\rho_{\boldsymbol{k}}^{\phantom{\ast}}}

\newcommand{\rhosk}{\rho^{\ast}_{\boldsymbol{k}}}

\begin{document}
\title{
Dynamics and rheology of active glasses}


\author{T.F.F.~Farage}
\author{J.M.~Brader}
\affiliation{Department of Physics, University of Fribourg, CH-1700 Fribourg, Switzerland}


\begin{abstract}
Within the framework of mode-coupling theory, we present a simple model for describing dense
assemblies of active (self-propelled) spherical colloidal particles. 
For isotropic suspensions, we demonstrate that the glass transition is shifted to higher 
volume fraction by the addition of activity, in agreement with recent Brownian 
dynamics simulations. 
Activity-induced changes in the static structure factor of the fluid are predicted.
The mechanical response of an active glass to applied strain is shown to be softer 
than the corresponding passive glass; both the nonergodicity parameter 
and the yield stress reduce with increasing activity. 
 
\end{abstract}


\maketitle
In recent years, considerable progress has been made in understanding the physical mechanisms 
underlying the collective motion of living matter, from macroscopic systems, such as fish schools or 
bird flocks, to the microscopic level, where the constituents are bacteria, cells or 
filaments in the cytoskeleton
\cite{collective_motion2012, ramaswamy2010, romanczuk2012, cates_bacteria2012}. 
The common feature among these disparate systems is that the constituents, be 
they birds or bacteria, exhibit active self-propulsion, whereby they are driven in a certain  
direction until reoriented either by a collision or some other physical process. 
The consequences of particle activity on the deformation and flow properties of fluid states, for 
which particle crowding is not too severe, have been addressed in a number of recent experimental 
\cite{rafai2010} and theoretical \cite{hatwalne2004, heidenreich2011, saintillan2010, wysocki2013} 
studies. 

Only very recently has attention been devoted to the dynamics of dense assemblies of 
self-propelled particles around the glass transition 
\cite{henkes2011, ni2013, wysocki2013,  bialke2012, berthier2013, berthier_disks_2013}. 
Within this high density regime there exists a nontrivial interplay between activity 
and the intrinsic slow structural relaxation arising from particle caging.  
With the exception of a spin-glass inspired study of 
schematic glassy models \cite{berthier2013},
all work to date has been based on computer simulations of 
interacting active disks or spheres. 
These studies have revealed several important generic features, such as shifted
glass \cite{ni2013, berthier2013, berthier_disks_2013} and crystallization transitions
\cite{bialke2012}.  
However, the high density behaviour of these simple model systems has yet to be described by microscopic theory and 
a unifying framework remains to be found.

In this Letter we propose a first-principles approach which captures some essential features of the collective dynamics 
in dense active suspensions, namely the dependence on activity of the glass transition, the 
static structure factor and the yield stress. 
Starting from a time-local, overdamped Langevin equation, we derive the corresponding Fokker-Planck 
equation for the time-evolution of the $N$-body probability distribution. 
Mode-coupling approximations then yield a closed and numerically tractable theory. 
In contrast to \cite{berthier2013}, which starts from a generalized Langevin equation, 
non-Markovian time evolution is an output of our treatment of the collective many-body dynamics. 

We consider a system of $N$ interacting, active, spherical Brownian particles with spatial coordinate 
$\rr_i$ and orientation specified by an embedded unit vector $\p_i$. 
Each particle experiences a self propulsion of speed $v_0$ in its direction of orientation and 
a one-body force generated by an external steady shear flow with velocity gradient
$(\nabla{\bf v})_{\alpha\beta}\equiv(\kapp)_{\alpha\beta}=\gam\delta_{\alpha x}\delta_{\beta y}$. 
Omitting hydrodynamic interactions the particle motion can be modelled by a pair 
of coupled stochastic differential (Langevin) equations
\begin{align}\label{full_langevin}
\dot{\rr}_i &= v_0\p_i + \kapp\cdot\rr_i - \zeta^{-1}\nab_{\rr_i}U_N + \xxi_i(t),
\notag\\
\dot{\p}_i &= \eeta_i\times\p_i + \boldsymbol{\omega}\times \p_i,
\end{align}
with potential energy $U_N$, friction coefficient $\zeta$, and angular velocity 
$\boldsymbol{\omega}\!=\!\nab_{\rr}\times\left( (\kapp\!-\!\kapp^{T})\!\cdot\!\rr \right)\!/2$.
%
The stochastic vectors $\xxi_i(t)$ and $\eeta_i(t)$ have zero mean and are delta correlated: 
$\langle\xxi_i(t)\xxi_j(t')\rangle=2D_t\boldsymbol{1}\delta_{ij}\delta(t-t')$ and 
$\langle\eeta_i(t)\eeta_j(t')\rangle=2D_r\boldsymbol{1}\delta_{ij}\delta(t-t')$. 
The rotational diffusion coefficient, $D_r$, is of thermal
origin and is thus related to the translational diffusion coefficient, $D_t=k_BT/\zeta$, according to $D_r=3D_t/d^2$, with particle diameter $d$.
In a description of run-and-tumble particles, 
the coefficient $D_r$ would determine the tumbling rate of the particles \cite{pototsky2012,cates_bacteria2012,cates2013}.




An alternative description of the dynamics is given by the configurational probability
distribution, 
$\Psi\!\equiv\!\Psi(\rr^N\!\!,\p^N\!\!,t)$ \cite{gardiner}, which evolves according to
\begin{equation}\label{smol_eq}
	\frac{\partial}{\partial t}\Psi = \Omega\,\Psi. 
\end{equation}
In the limit $|\,\boldsymbol{\omega}|/D_r\!\ll\!1$, which is appropriate for the low shear rates to be considered in this work, the 
dynamical (Smoluchowski) operator is given by
\begin{align}\label{smol_op}
	\Omega = \sum_{i=1}^{N}&\Big(\nab_{\rr_i}\cdot\left(D_t\left(\nab_{\rr_i}-\beta\F_i\right)\right)
	\notag\\
	&\hspace*{-0.18cm}	-\nab_{\rr_i}\cdot\left(v_0\,\p_i+\kapp\cdot\rr_i\right) + D_r\R_i^2\Big),
\end{align}
where $\beta\!\equiv\!(k_BT)^{-1}$, $\F\!=\!-\nab_{\rr}U_N$ is the force due to potential interactions and 
$\R\equiv(\p\times\nab_{\p})$ accounts for rotational diffusion. 
The term $-\nab_{\rr_i}\!\!\!\cdot \!v_0\p_i$ expresses the essential coupling between
translational and rotational motion.

Before addressing the full dynamics \eqref{smol_eq}, we first focus our attention on the simpler
case of 
a single particle subject only to rotational diffusion and self propulsion. 
The distribution for a single particle undergoing only these two types of motion evolves according to
\begin{align}\label{single}
\frac{\partial\Psi(\rr,\p,t)}{\partial t}=\left(-\nab_{\rr}\cdot v_0\,\p + D_r\R^2\right)\Psi(\rr,\p,t). 
\end{align}
On length-scales large compared to the persistence length of the particle trajectory, and under the 
condition of constant $v_0$ and $D_r$, the operator on the r.h.s. of \eqref{single} can be 
well approximated by a random-walk with diffusivity 
$D_{\text{eff}}=v_0^2\,/D_r$ \cite{othmer1988, cates_bacteria2012, cates2013}
\begin{align}
\left(-\nab_{\rr}\cdot
v_0\,\p+D_r\R^2\right)\,\rightarrow\,
D_{\text{eff}}\nab_{\rr}^{2}.  
\end{align}
Applying this approximation to the full dynamics \eqref{smol_eq}, we obtain an effective
Smoluchowski operator,
\begin{equation}\label{eff_smol_op}
	\Omega_{\text{eff}} =\sum_{i=1}^{N}\Big(\nab_{\rr_i}
	\cdot D_t\left(\alpha\nab_{\rr_i}
	-\beta\F_i\right)
	-\nab_{\rr_i}\cdot\kapp\cdot\rr_i\Big),
\end{equation}
where $\alpha\equiv(1+(D_{\text{eff}}/D_t))$. 
The approximate dynamics specified by \eqref{eff_smol_op} makes clear that 
particle activity does not simply renormalize the translational 
diffusion coefficient, but rather modifies fundamentally the balance between thermal and potential forces. 
An intuitive consequence of this shifted balance is that activity will influence the location of the glass transition. 

A convenient probe of the collective dynamics is provided by the density correlator
\begin{align}\label{density_correlator}
\Phik(t)\equiv\frac{1}{NS_k}\langle\rhosk
e^{\omadj_{\text{eff}}t}\rhok\rangle,
\end{align}
where 
$\rhok=\sum_{i}e^{i\kk\cdot\rr_i}$ is the
Fourier transform of the density, $S_k$ is the static structure factor, and $\langle\cdot\rangle$ 
denotes an equilibrium average. The adjoint Smoluchowski operator is given by
\begin{equation}\label{adj_smol_op}
	\omadj_{\text{eff}} =
\sum_{i=1}^{N}\Big(D_t\left(\alpha\nab_{\rr_i}+\beta\F_i\right)\cdot\nab_{\rr_i} +
\kapp\cdot\rr_i\cdot\nab_{\rr_i}\Big), 
\end{equation}
where for arbitrary functions $A$ and $B$ the adjoint is defined according to $\int d\rr^N\! A\,\om
B=\int d\rr^N\! B\,\omadj A$.

\begin{figure}[ht]
\begin{center}
  \subfigure{\epsfig{figure=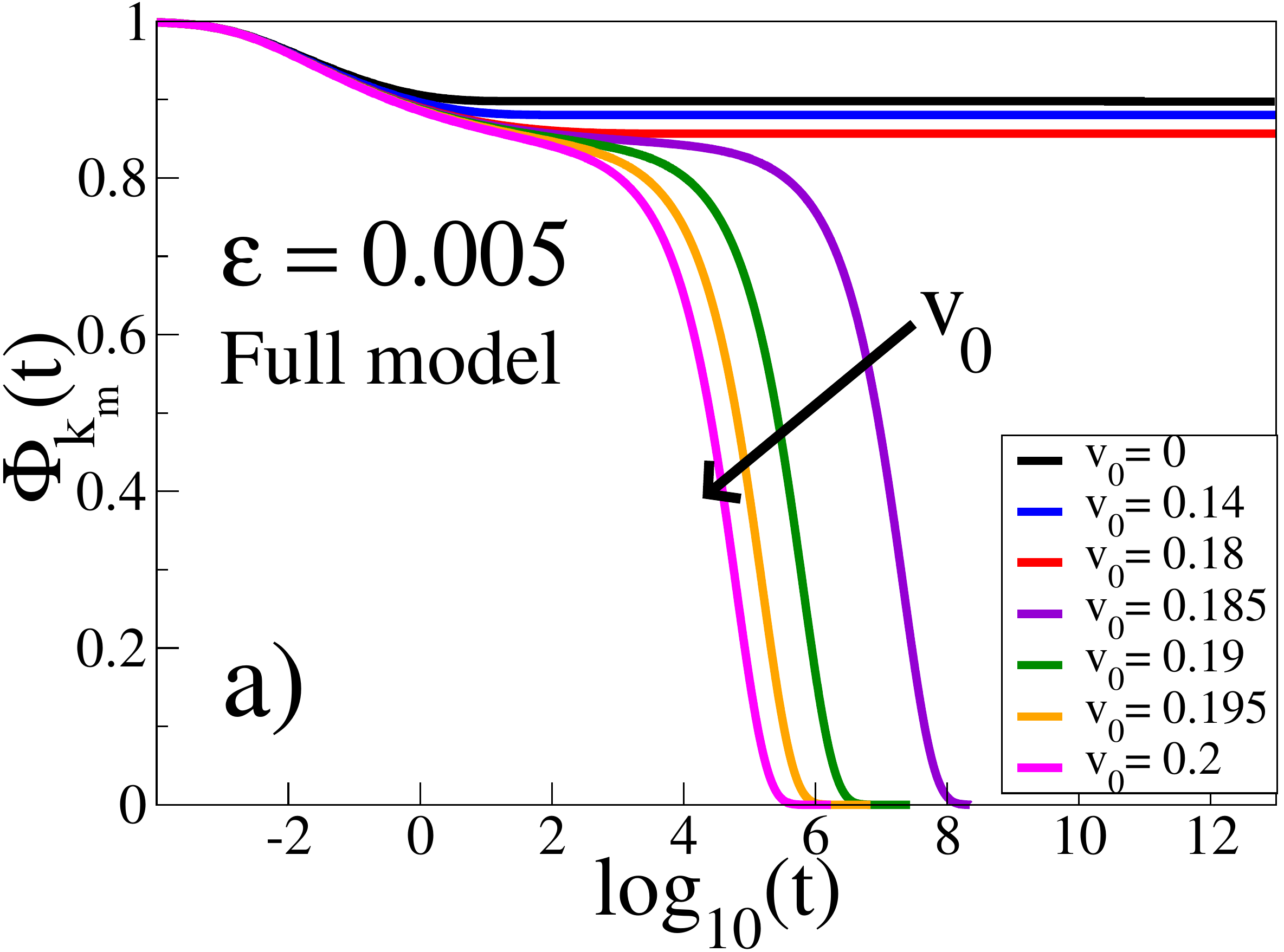,width=0.485\linewidth}}
  \subfigure{\epsfig{figure=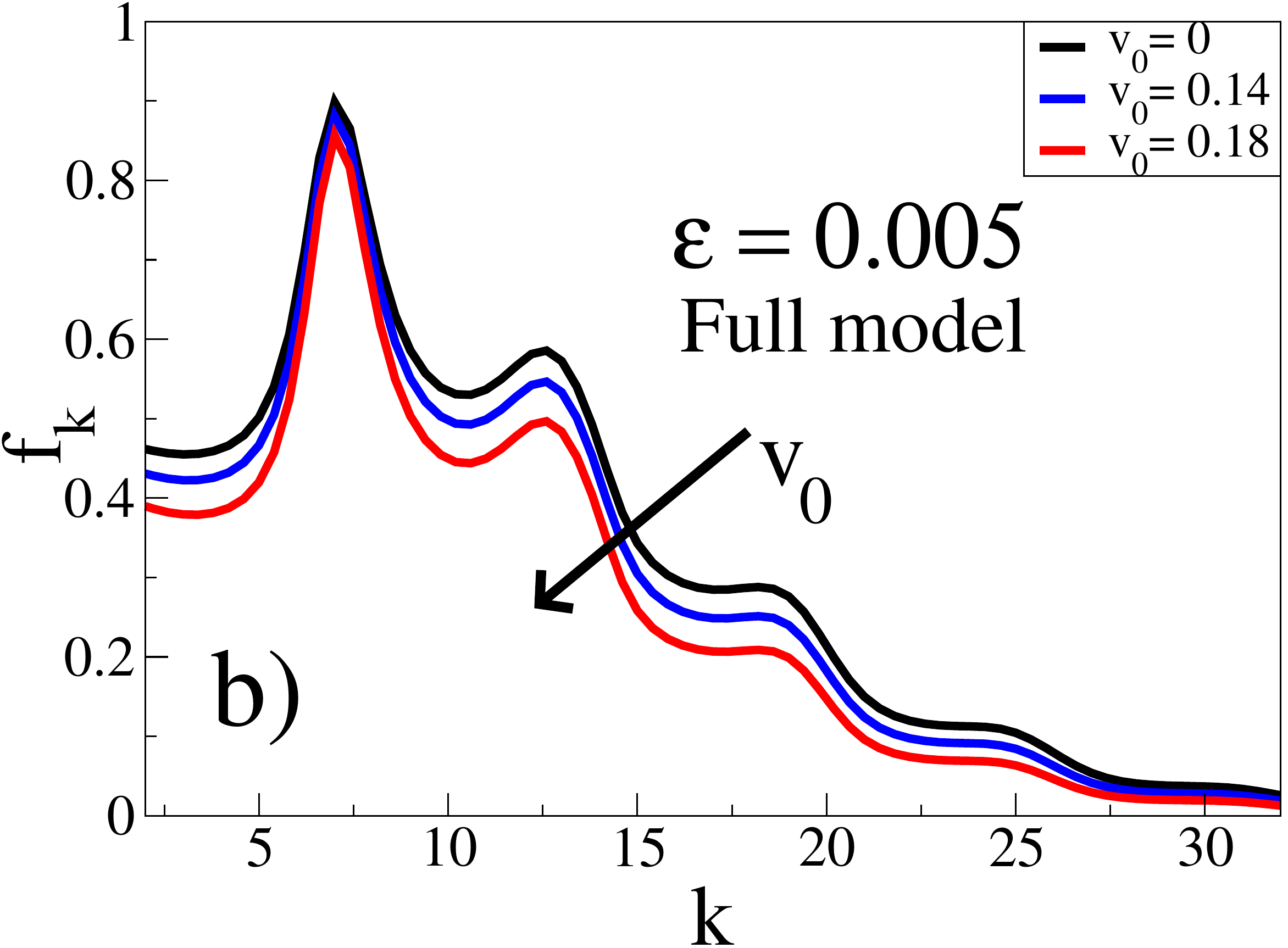,width=0.485\linewidth}}\\
\end{center}
\vspace*{-0.7cm}
\begin{center}
  \subfigure{\epsfig{figure=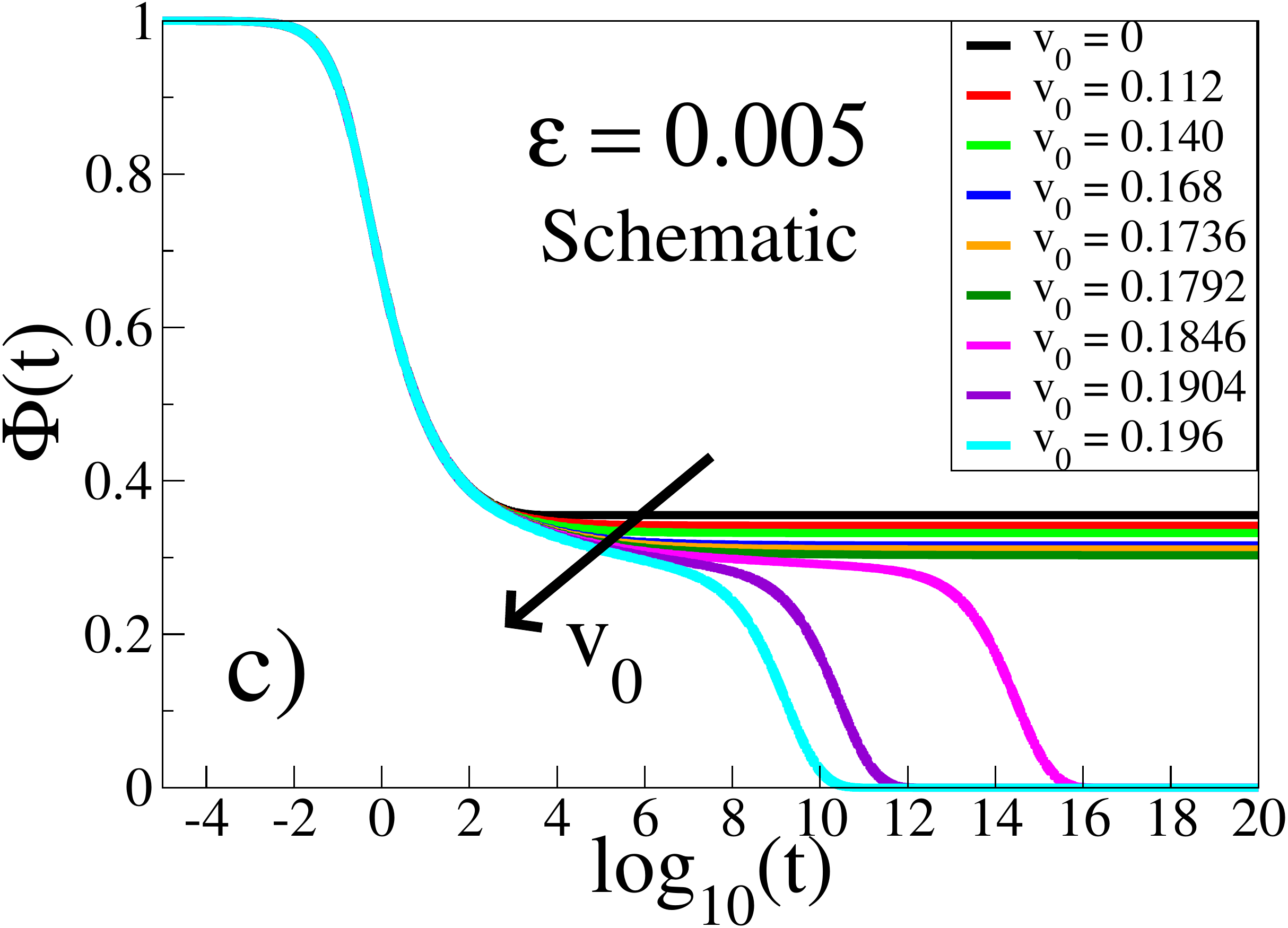,width=0.485\linewidth}}
  \subfigure{\epsfig{figure=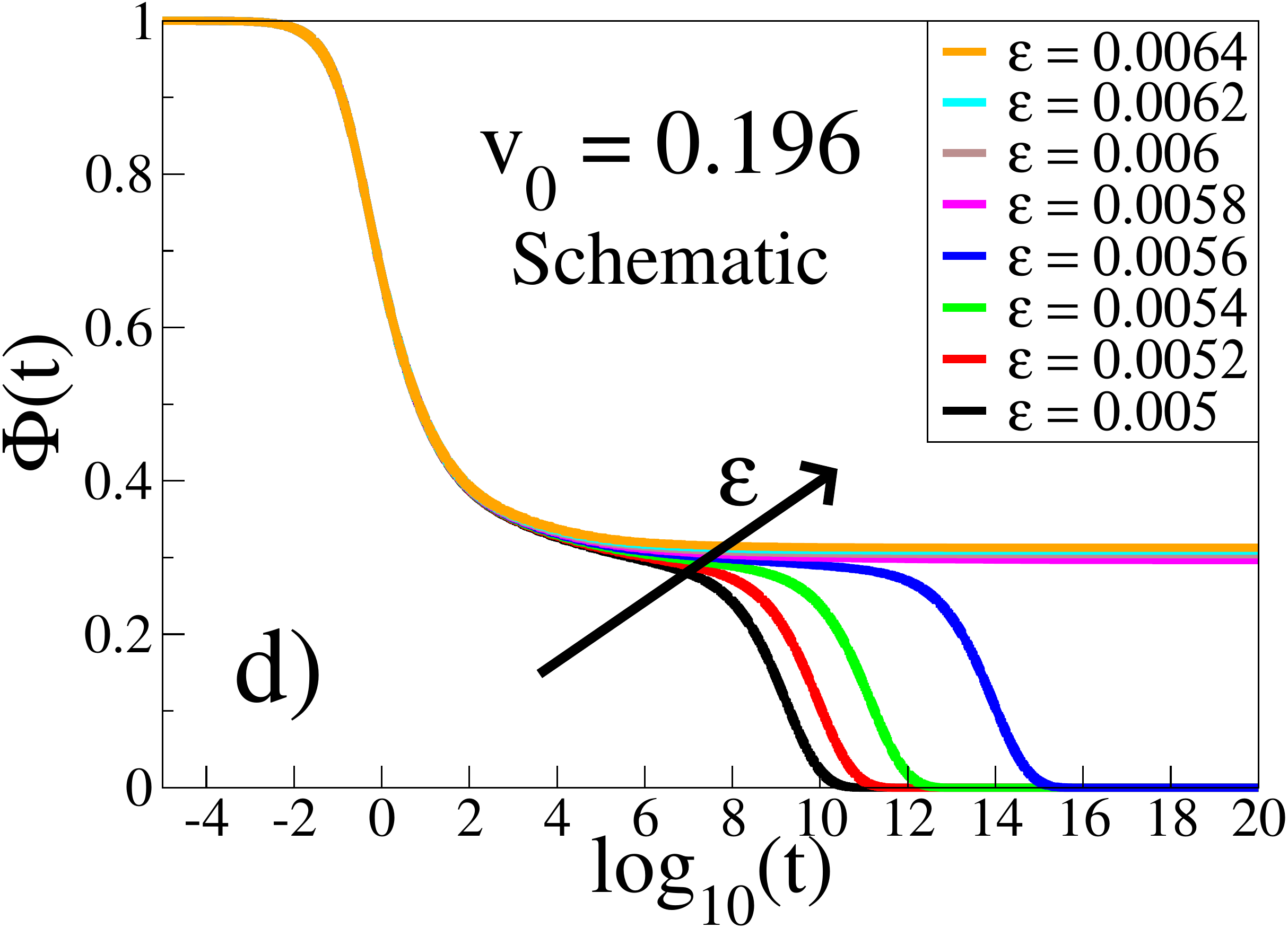,width=0.485\linewidth}}\\
\end{center}
\vspace{-0.5cm}
\caption{{\bf (a)} Melting the glass by activity: decay of the correlation function 
at fixed $\varepsilon = 0.005$ and $k_{\rm m}=7$ for increasing $v_0$. 
{\bf (b)} Activity softens the glass: nonergodicity parameter as a function of $k$ for increasing $v_0$.
{\bf (c)} Melting the glass by activity: schematic correlation functions calculated using Eq.\eqref{schem_eom}. 
{\bf (d)} Revitrifying the active glass: increasing the coupling for a fixed $v_0=0.196$, schematic correlation functions.
}
\label{fig_correl}
\end{figure}
  
The relatively simple form of the adjoint effective operator \eqref{adj_smol_op} facilitates 
the application of mode-coupling methods to approximate the density correlator 
\eqref{density_correlator}. 
Starting from a well known Zwanzig identity and using the projection 
operator formalism \cite{zwanzig2001}, we obtain a non-Markovian equation of motion for
the correlation function of active particles in the absence of shear
\begin{align}\label{eom}
	&\dot{\Phi}_{\boldsymbol{k}}(t)\!+\! A_{\boldsymbol{k}}\Gammak 
\left\{\Phik(t)\!+\!\frac{1}{A_{\boldsymbol{k}}^2}\int_{0}^{t}\!\!dt'\mk(t\!-\!t')
	\dot{\Phi}_{\boldsymbol{k}}(t')\right\} = 0, 
\end{align}
where $A_{\boldsymbol{k}}\equiv 1\!+\!S_kv_0^2(D_tD_r)^{-1}$ and 
$\Gamma_{\boldsymbol{k}}=D_tk^2/S_k$. The time-translationally invariant memory kernel is given by
\begin{equation}\label{memory}
	\mk(t)=\frac{\rho}{16\pi^3}\int d\qq\frac{S_kS_qS_p}{k^4}\mathcal{V}^2_{\kk\p\qq}
	\Phi_{\boldsymbol{q}}(t)\Phi_{\boldsymbol{p}}(t),
\end{equation}
with $\p=\kk-\qq$ and bulk number density $\rho$. 
The vertex function is given by $\mathcal{V}_{\kk\p\qq}=\kk\cdot(\qq c_{\qq} + \p c_{\p})$, 
where $c_{\kk}=(1- S_k^{-1})/\rho$ is the direct correlation function.

The active mode-coupling equation (\ref{eom}) is a central result of this Letter.  
The solution exhibits a bifurcation at sufficiently high coupling (density or attraction strength) 
accounting for dynamic arrest. 
Eqs.\eqref{eom} and \eqref{memory} form a closed theory for the density correlator, which enables
the competition between structural relaxation and activity to be investigated. 
The only required input quantities are $\rho$, $v_0$ and $S_{k}$. 
The long-time limiting solution of \eqref{eom} can be obtained by setting the expression in curly 
brackets equal to zero, $\{\cdot\}=0$. 
The modified balance between thermal and interaction forces in the many-body expression
\eqref{eff_smol_op} 
(without shear) is manifest in our approximate equation \eqref{eom} via the activity
dependence 
of $A_{\boldsymbol k}$.

\begin{figure}[t]
\hspace*{-0.5cm}
\includegraphics[width=0.43\textwidth]{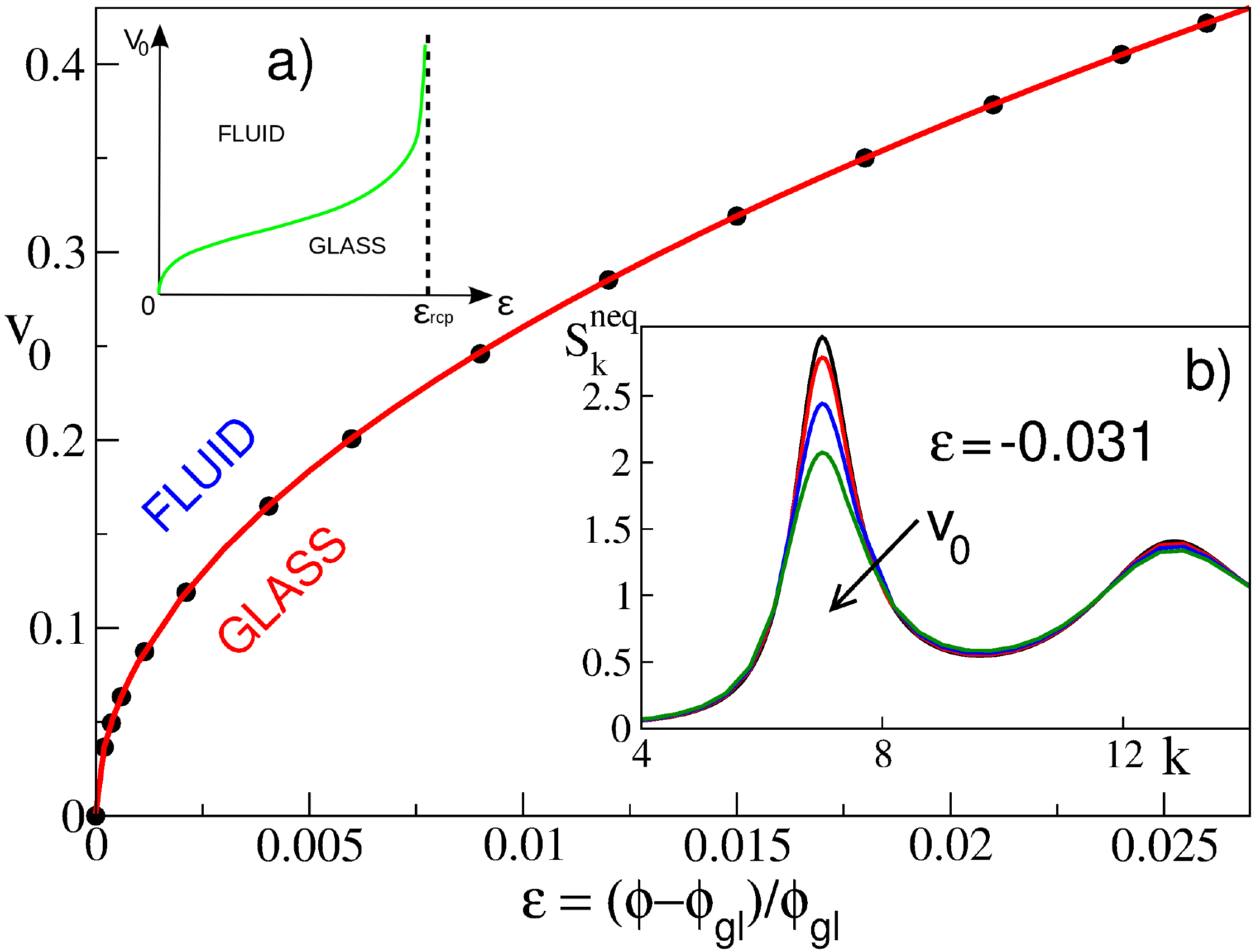}
\caption{Phase diagram of active particles. Points are obtained from numerical 
solution of \eqref{eom} and the (red) line is a one parameter fit using the schematic 
Eq.\eqref{schem_eom}.  
{\bf (a)}: hand-drawn sketch of a speculative phase boundary for volume fractions between
the glass transition ($\varepsilon=0$) and random close packing ($\varepsilon_{\rm rcp}$). 
{\bf (b)}: nonequilibrium structure factor for $v_0=0,\,0.05,\,0.1$ and $0.15$, calculated 
using Eq.\eqref{mct_structure}.  
} 
\label{fig_phase_diagram}
\end{figure}

We have solved Eqs.\eqref{eom} and \eqref{memory} numerically for hard-spheres using Percus-Yevick static 
structure factors as input and setting $D_t\!=\!1$. The numerical discretization is 
identical to that employed in \cite{franosch1997} and predicts for the passive system a glass
transition at volume 
fraction $\phi_{\rm gl}\equiv \rho_{\rm gl}\pi d^3/6=0.515912$. 
In Fig.\ref{fig_correl}a we show the time-evolution of correlators evaluated at approximately the first peak 
in $S_k$, $k_{\rm m}\!=\!7$, for a volume fraction in the glass, $\phi\!=\!0.5185$, 
corresponding to a separation parameter $\varepsilon\equiv(\phi-\phi_{\rm gl})/\phi_{\rm gl}=0.005$. 
For small activity values, the correlator does not decay and the system remains arrested. 
However, the decrease in plateau height, which is related to the elastic constant, indicates a softening of the glass. 
Beyond a critical value $v_0^*=0.1836$, the activity is sufficent to fluidize the glass and the
correlator relaxes. 
Fig.\ref{fig_phase_diagram} shows the locus of points in the ($v_0,\phi$) plane separating 
fluid and glassy states: activity clearly shifts the transition to higher $\phi$ values. 
We note that recent simulations of self-propelled Brownian disks have shown 
that crystallization is shifted to higher volume fraction as activity is increased \cite{bialke2012}, 
possibly indicating that common mechanisms underly the processes of vitrification and crystallization in active systems. 
In Fig.\ref{fig_correl}b we show the nonergodicity parameter, $f_k\equiv\Phi_k(t\!\rightarrow\!\infty)$, for several 
values of $v_0$. 
The activity-induced softening of the glass is a nontrivial function of $k$ and is found to be weakest for values 
around the main peak of the equilibrium static structure factor.

The nonequilibrium structure factor of the active system can be addressed using the method of 
integration through transients (ITT) \cite{fuchs_cates2002}. 
Treating the $v_0$ dependent contribution to the effective operator \eqref{eff_smol_op} (in the
absence of shear) as an inhomogeneity, 
and then formally solving \eqref{smol_eq}, generates a Green-Kubo-type formula for steady-state averages
\begin{align}\label{GK}
\langle f\,\rangle_{\rm neq}\! =\! \langle f\,\rangle +\! \int_0^{\infty}\!\!\!dt' \sum_i \beta D_{\rm eff}\left\langle\!( \beta\,{\bf F}_i^2 
+ \nabla_{\boldsymbol r_i}\cdot{\bf F}_i )e^{\Omega^{\dagger}_{\rm eff}t'}\!f \!\right\rangle,
\end{align}
where $f$ is an arbitrary function. Applying this result to calculate 
$\langle\rho^*_{\boldsymbol k}\rho^{}_{\boldsymbol k}\rangle_{\rm neq}$ and then employing 
mode-coupling projection operator approximations yields a simple result for the nonequilibrium structure factor
\begin{align}\label{mct_structure}
S_k^{\rm neq} = S_k \;+\; \frac{1}{2}D_{\rm eff} k^2(1-S_k)\!\int_0^{‌\infty}\!\!dt\,\Phi_k(t)^2, 
\end{align}
where $\Phi_k(t)$ is a solution of \eqref{eom}. 
In Fig.\ref{fig_phase_diagram}b we show numerical solutions of \eqref{mct_structure} for $\phi\!=\!0.5$ 
($\varepsilon\!=\!-0.031$) and several values of $v_0$. 
We predict that the structure of active hard spheres differs from that of the passive 
system, particularly in the vicinity of the main peak, the height of which decreases with increasing $v_0$. 
This finding compares favourably with the simulations of Ni {\it et al.} \cite{ni2013}, who observed 
similar behaviour close to random close packing. 
In view of the Hansen-Verlet freezing criterion, which states that crystallization should occur when the main 
peak of $S_{k}$ exceeds the value $2.85$, a reduction of peak height with increasing $v_0$ is consistent 
with the shifted phase boundary reported by Bialk\'e {\it et al.} \cite{bialke2012}.

Experience with the passive mode-coupling theory has shown that wavevector dependent equations can be 
exploited to develop ``schematic" models, which simplify the equations while retaining the essential physics. 
This approach has proven particularly useful when addressing sheared systems \cite{brader_pnas2009}, for 
which spatial anisotropy complicates the numerical solution of the equations. 
Schematic models are obtained by suppressing wavevector indices, resulting in a single-mode description. 
Applying this strategy to \eqref{eom} yields a schematic equation of motion for (unsheared) 
active particles
\begin{equation}\label{schem_eom}
	\dot{\Phi}(t)+ A
	\left\{\Phi(t)+\frac{1}{A^2}\int_{0}^{t}dt'm(t-t')
	\dot{\Phi}(t')\right\} = 0,
\end{equation}
where $m(t)=\nu_1\Phi(t) + \nu_2\Phi^2(t)$, with
$\nu_1=2(\sqrt{2}-1)+\varepsilon/(\sqrt{2}-1)$ and $\nu_2=2$. 
This form of the memory function, as well as the relation between $\nu_1$ and 
$\nu_2$, is taken from the established $F_{12}$ schematic model \cite{gotze_book}, which reproduces 
essential features of the passive theory. Activity enters \eqref{schem_eom} 
via $A\equiv 1\!+\!\nu v_0^2$, where the parameter $\nu$ mimics the role of the static structure 
factor appearing in $A_{\boldsymbol k}$. 
We choose the value $\nu=0.12755$, which ensures that the phase boundaries of the full \eqref{eom} 
and schematic \eqref{schem_eom} theories match as closely as possible (see Fig.\ref{fig_phase_diagram}). 

In Fig.\ref{fig_correl}c, we show the schematic density correlator 
for a glassy state, $\varepsilon\!=\!0.005$, at different values of $v_0$. 
The qualitative behaviour is close to that found from the full theory: 
increasing $v_0$ first diminishes the height of the plateau and then, when a 
critical value ($v_0^{*}\approx0.182$) is exceeded, the system relaxes as activity 
melts the glass. As $v_0$ is further increased the relaxation time continues to 
decrease, eventually saturating at very high activities. 
In Fig.\ref{fig_correl}d, we illustrate how a glassy state can be recovered from an active fluid 
by increasing $\varepsilon$. Choosing $v_0=0.196$, we find that a glass transition occurs at a 
critical $\varepsilon^{*}\approx0.0057$, which lies above the passive value $\varepsilon=0$. 
The schematic phase boundary shown in Fig.\ref{fig_phase_diagram} (red line) summarizes the
parameter space 
for which we obtain glassy solutions. 
Neither \eqref{eom} nor \eqref{schem_eom} contain information about random close packing 
(at around $\phi_{\text{rcp}}\approx0.64$). 
Consequently, the predictions of our theory will become unreliable as the volume fraction 
approaches $\phi_{\text{rcp}}$. 
In Fig.\ref{fig_phase_diagram}a we sketch a plausible form of the phase boundary for volume fractions 
from $\phi_{\rm gl}$ to $\phi_{\text{rcp}}$, which may be expected from a more complete theory. 

As a further application of our theory, we address the rheology of active glasses. 
Both passive particles under shear and unsheared 
active particles represent nonequilibrium systems. However, the nature of the driving forces 
is fundamentally different. The former is a ``coherent" driving, acting globally to melt the glass 
for any finite $\dot\gamma$ value, whereas the latter is ``incoherent", acting on the particle level, 
which permits the existence of active glasses for finite $v_0$. 
A further important distinction is that shear breaks the symmetry of the system, such that the density 
correlator becomes anisotropic. 
This anisotropy complicates considerably the numerical solution of mode coupling equations and this 
has provided strong motivation for the development of schematic models \cite{brader_pnas2009}.

The shear stress, $\sigma_{\rm xy}$, of a passive glass tends to a finite value as $\dot\gamma$ is reduced adiabatically 
towards zero. 
Within mode-coupling theory, this dynamical yield stress,
$\sigma_{\text{yield}}\equiv\sigma_{xy}(\gam\rightarrow0)$, emerges discontinously 
upon entering the glass and then increases with density 
according to a power law $\sigma_{\rm yield}\sim\varepsilon^{\frac{1}{2}}$. 
It is known that the phenomenology of the wavevector dependent mode-coupling theory 
of sheared suspensions \cite{brader_prl2008} can be well represented for steady shear by the 
schematic F$_{12}^{\gam}$ model \cite{fuchs_cates2003}. 
Incorporating a similar treatment of shear into our schematic equation \eqref{schem_eom} is straightforward and 
simply requires that we make the replacement 
\begin{align}\label{replacement}
m(t)\rightarrow m(t;\gam)=\frac{\nu_1\Phi(t) + \nu_2\Phi^2(t)}{1+(\gam t)^2}.
\end{align}
We can now use this modified schematic model to investigate how the yield stress is influenced by 
activity. 

Following \cite{brader_pre2012} and applying the integration through transients methods, 
it is straightforward to generate from (\ref{adj_smol_op}) an exact generalized Green-Kubo relation 
for the stress. 
Mode-coupling approximations to this expression enable the following schematic expression to be inferred
\begin{equation}\label{schem_GK}
	\sigma_{xy}=\gam\int_{0}^{\infty}\!\!dt\,\Phi^2(t),
\end{equation}
where $\Phi(t)$ is solution of (\ref{schem_eom}), with memory function \eqref{replacement}.
%
%
\begin{figure}[t]
\begin{center}
\includegraphics[width=0.46\textwidth]{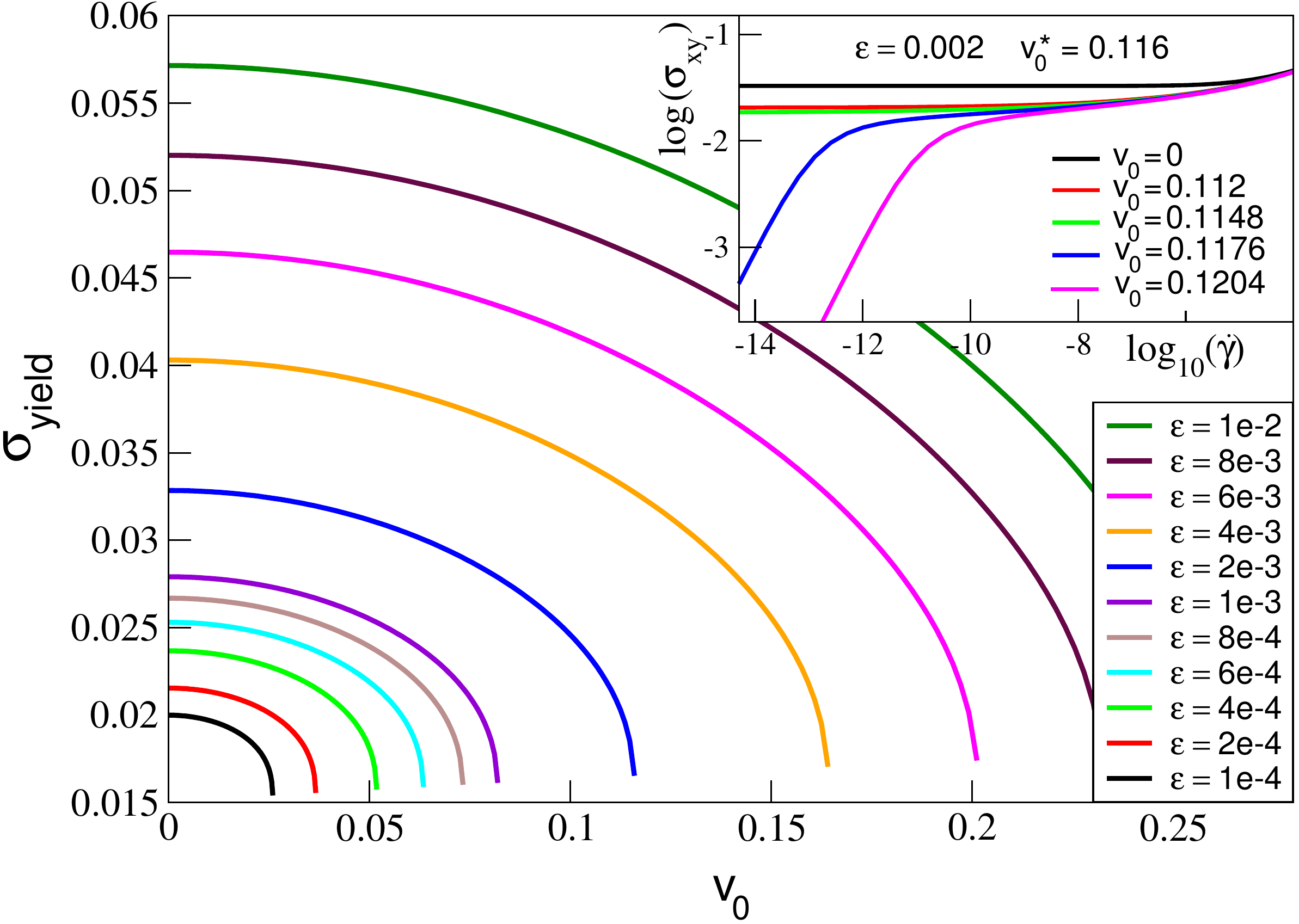}
\vspace*{-0.2cm}
\caption{Yield stress $\sigma_{\text{yield}}$ calculated using Eq.\eqref{schem_GK} as a function 
of $v_0$ for several values of $\varepsilon$. 
Inset: flow curves at $\varepsilon=0.002$ for velocities around the critical value $v_0^{*}=0.116$. 
}
\label{fig_yield}
\end{center}
\end{figure}
%
In Fig.\ref{fig_yield}, we show numerical results for $\sigma_{\text{yield}}$ as a function of
the $v_0$ for several values of $\varepsilon$. 
The inset shows a selection of flow curves ($\sigma_{\rm xy}$ as a function of $\dot\gamma$) for 
a glassy state, $\varepsilon=0.002$, around the corresponding critical velocity $v_0^*=0.116$. 
As $\dot\gamma\!\rightarrow\!0$ the curves for $v_0<v_0^*$ tend to a plateau, whereas for $v_0>v_0^*$ 
they enter a Newtonian regime. 
For all $\varepsilon$ values the yield stress decreases significantly with increasing $v_0$: 
Active glasses can thus be softened, without melting, by an increase in the activity. 
For a given $\varepsilon$, the yield stress jumps discontinuously to zero as $v_0$ exceeds its
value 
at the phase boundary (see Fig.\ref{fig_phase_diagram}), this is where the curves end in
Fig.\ref{fig_yield}. 
We find that these minimum values of $\sigma_{\text{yield}}$ increase only slightly with $\varepsilon$, 
thus raising the question of the existence of a universal yield stress value at the phase boundary. 
Finally, despite the similarity of the curves shown in Fig.\ref{fig_yield} for different $\varepsilon$ 
values, scaling $\sigma_{\rm xy}$ and $v_0$ does not cause the yield stress to collapse onto a master 
curve. 
 
In conclusion, we have derived from first-principles a mode-coupling theory describing the competition 
between slow structural relaxation and particle activity. 
This microscopic theory makes parameter-free predictions for the fluid-glass phase boundary, the density 
correlator and nonequilibrium static structure factor. 
Inspired by these results, we have developed a simplified schematic model, which enables the study
of sheared 
active glasses. 
We find that activity softens the glass by reducing the correlator plateau value, thus reducing the yield stress.  
In view of very recent experimental progress in the design of self-propelled particles whose
propulsion can be controlled by blue light \cite{palacci2013}, the phenomenon of active
glass softening could offer new perspectives in the design of amorphous solids.


\bibliography{references.bib}
\newpage
\end{document}